\documentclass[reprint,amsmath,amssymb,aps,prl,superscriptaddress]{revtex4-1}

\usepackage{graphicx}
\usepackage{dcolumn}
\usepackage{bm}
\usepackage{hyperref}
\usepackage{xurl}
\hypersetup{breaklinks=true}

\usepackage{mathtools}
\hypersetup{colorlinks=true, linktoc=all, linkcolor=blue, linktocpage, citecolor=blue}

\begin{document}

\title{Composition Dependent Instabilities in Mixtures With Many Components}

\author{Filipe C. Thewes}
%\email{filipe.cunhathewes@uni-goettingen.de}
\affiliation{%
 Institut für Theoretische Physik, Georg-August-Universität Göttingen, 37077 Göttingen, Germany
}%
\author{Matthias Krüger}
\affiliation{%
 Institut für Theoretische Physik, Georg-August-Universität Göttingen, 37077 Göttingen, Germany
}%
\author{Peter Sollich}%
 %\email{}
\affiliation{%
 Institut für Theoretische Physik, Georg-August-Universität Göttingen, 37077 Göttingen, Germany
}%
\affiliation{King's College London, Department of Mathematics, Strand, London WC2R 2LS, U.K.\\}

\date{\today}

\begin{abstract}
    Understanding the phase behavior of mixtures with many components is important in many contexts, including as a key step toward a physics-based description of intracellular compartmentalization. Here, we study the instabilities of a mixture model where the second virial coefficients are taken as random Gaussian variables. Using tools from free probability theory we obtain the exact spinodal curve and the nature of instabilities for a mixture with an arbitrary composition, thus lifting the assumption of uniform mixture component densities pervading previous studies. We show that, by controlling the volume fraction of only a few components, one can systematically change the nature of the spinodal instability and achieve demixing for realistic scenarios by a strong {\em composition imbalance amplification}. This results from a non-trivial interplay of entropic effects due to non-uniform composition and complexity in the interactions. Our approach allows for the inclusion of any finite number of structured interactions, leading to a competition between different forms of demixing as density is varied.
\end{abstract}

\maketitle

Phase separation is an important phenomenon and is especially rich in mixtures with many components. In particular, biological mixtures such as the cytoplasm show a complex phase behavior believed to be a key driver in the formation of nucleoli and other intracellular structures~\cite{Alberti2017,Lafontaine2020, Sear2007, Berry2018,Choi2020}. 
Typically, these systems demix into liquid droplets with different compositions, where each phase is enriched in a number of components and depleted in others~\cite{Berry2018,Keating2012}. 

Within a mean-field approach, previous studies have explored the rich behavior of complex mixtures by investigating the number of phases formed for appropriately tuned~\cite{Jacobs2021}, evolutionarily optimized~\cite{Zwicker2022} and random~\cite{Jacobs2013,Jacobs2017} interactions. Within the latter approach~\cite{Sear2003} dynamical properties have also been investigated, including early time instabilities using random matrix theory~\cite{Sear2003} and direct numerical simulations~\cite{shrinivas2021} of continuum field theories (model B).

Although these studies have been able to illuminate some of the behavior of complex mixtures, a key restrictive assumption common to all of them is uniform composition, meaning all species are present in the system in equal amounts. However, biological mixtures rarely satisfy this condition~\cite{Sear2005,Gasic2021,neidhardt1987}, and the size of intracellular structures is in fact heavily dependent on the composition of the cytoplasmic pool~\cite{Weber2015,Goehring2012}. In addition, experimental protocols often rely on controlling the composition of the mixture to study its phase behavior~\cite{Gasic2021,Riback2020,Zhang2015,Fritsch2021}. 

With this in mind, our aim in this letter is to open the door to exploring the full composition-dependent complexity of multi-component mixtures. To achieve this, we build on the model of Sear and Cuesta~\cite{Sear2003} but lift the drastic simplification of uniform composition. We show that systematically changing the number density of only a few components enables one to control the nature of the instabilities and, as a consequence, the phases that can be formed.

% Model
%%%%%%%%%%%%%%%%%%%%%%%%%%%%%%%%%%%%%%%%%%%%%%%%%%%%%%

\textit{Model and general results}: Following~\cite{Sear2003} we study a mixture of $M$ different components, labeled by Greek letters,  with interactions described by the second virial coefficients $\epsilon_{\alpha\gamma}$. The mean-field free energy density $f$ is given by~\cite{Weber2019}
\begin{equation}
        f = \frac{1}{2}\sum_{\alpha,\gamma=1}^M \rho_\alpha \epsilon_{\alpha\gamma}\rho_\gamma + T\sum_{\alpha=1}^M \rho_\alpha\ln\rho_\alpha + T\rho_0\ln\rho_0
        \label{eq:freeEnergy}
\end{equation}
where $\rho_\alpha=N_\alpha/V$ is the number density of species $\alpha$, $T$ is temperature and we use $k_{\rm B}=1$. The last term in~(\ref{eq:freeEnergy}) is the entropic contribution of an implicit solvent, interacting only via volume exclusion. We define the total density as $\rho=\sum_{\alpha} \rho_\alpha$, the average density per component as $\bar\rho = \rho/M$, and work with units such that $\rho_0=1-\rho\geq 0$.

From~(\ref{eq:freeEnergy}) we obtain the $M\times M$ Hessian matrix $H_{\alpha\gamma}=\partial^2f/(\partial \rho_\alpha\partial\rho_\gamma)$ as $\boldsymbol{H} =(T/\rho_0) \boldsymbol{uu}^\mathsf{T} + (T/\bar\rho)\textrm{diag}\left (1/y_\alpha \right ) + \boldsymbol{\epsilon}$ with $\boldsymbol{u}=(1,1\dots,1)^\mathsf{T}$ the constant vector and $\textrm{diag}(1/y_\alpha)$ a diagonal matrix with entries determined by the {\em relative densities} $y_\alpha=\rho_\alpha/\bar\rho$; by definition the latter have average $M^{-1}\sum_\alpha y_\alpha=1$.  Thermodynamic stability requires all eigenvalues of $\boldsymbol H$ to be non-negative. Otherwise, i.e.\ if the lowest eigenvalue $\lambda_{\min}$ is negative, the system is unstable to phase separation by spinodal decomposition.

The phase diagram in the ($\rho$,$T$)-plane for a fixed composition $\{y_\alpha\}$ splits into stable and unstable regions, separated by a spinodal line determined by the condition $\lambda_{\min}=0$~\cite{Sollich2002}. The nature of the spinodal instability is determined by the eigenvector $\boldsymbol{v}$ corresponding to $\lambda_{\min}$. We will be interested in instabilities of {\em condensation} type ($\boldsymbol v \sim \boldsymbol u$), where the densities of all species change by similar amounts, and of {\em demixing} type ($\boldsymbol v^\mathsf{T} \boldsymbol u\approx 0$), where some species are enhanced while others are depleted. We will show that the demixing case can be further split into delocalized or {\em random}, where all components of $\boldsymbol v$ are of similar order, and {\em localized} where a few species have much larger entries in $\boldsymbol v$ and thus dominate the demixing. 

Following Ref.~\cite{Sear2003}, we model the second virial coefficients $\boldsymbol \epsilon_{\alpha\gamma}$ as Gaussian random variables of mean $-b$ and variance $s^2$, drawn independently except for the symmetry constraint $\epsilon_{\alpha\gamma}=\epsilon_{\gamma\alpha}$. The Hessian matrix then reads
    \begin{align}
    \begin{split}
        \boldsymbol H &= \boldsymbol R_1 + \boldsymbol D + s\boldsymbol \eta\qquad \mbox{with}\\
        %\mbox{with}\ 
        \boldsymbol R_1 &= \left (-b + \frac{T}{\rho_0} \right ) \boldsymbol{uu}^\mathsf{T}, \quad
        \boldsymbol D = \frac{T}{\bar\rho}\textrm{diag}\left (\frac{1}{y_\alpha} \right )
        \label{eq:hessian}
    \end{split}
    \end{align}
and $\boldsymbol\eta$ a Wigner matrix with entries of zero mean and unit variance~\cite{akemann2011oxford}. Understanding thermodynamic instabilities then requires us to obtain the eigenvalue distribution or {\em spectrum} of $\boldsymbol H$, and specifically its lower edge $\lambda_{\min}$; we focus throughout on the interesting multi-component limit $M\gg 1$. 
    
The $s\boldsymbol\eta$ term in $\boldsymbol H$ produces a  continuous spectrum of eigenvalues, and this extends to $\boldsymbol D+s\boldsymbol \eta$ \footnote{Only if the entries of $\boldsymbol D$ are relatively close to each other. Otherwise, the spectrum may split intro multiple bulk pieces.}. The first term $\boldsymbol R_1$ can be viewed as a rank one perturbation; due to the so-called interlacing property of eigenvalues~\cite{hwang2004}, the spectrum of $\boldsymbol H$ for large $M$ is then either the same as the spectrum of $\boldsymbol D+s\boldsymbol \eta$, or $\boldsymbol R_1$ may give rise to a single outlier~\cite{Benaych-Georges2011}, which separates from the continuous bulk spectrum of eigenvalues. We therefore have two regimes: if an outlier exists to the left of the bulk, then it is the lowest eigenvalue $\lambda_{\min}$. Otherwise the lowest eigenvalue is given by the lower edge of the bulk itself.
    
Free probability~\cite{voiculescu1992,akemann2011oxford} is a powerful tool to obtain the statistics of eigenvalues and eigenvectors of large random matrices, provided they obey the so-called freeness criteria. A key insight is that freeness generically holds between $\boldsymbol D$, $s\boldsymbol\eta$ and $\boldsymbol R_1$ (see Supplemental Material at~\cite{} for discussion). We can thus use free probability to analyze the spectrum of the scaled Hessian $\boldsymbol H/M$; adopting also the scaling $s=M^{1/2}\tilde s$~\cite{Sear2003} ensures that all matrices involved have eigenvalues of $O(1)$. We find~(see Supplemental Material at~\cite{} for this and subsequent derivations) for the spinodal equation, which determines where $\lambda_{\min}=0$,
    \begin{equation}
    \psi(-\rho \tilde s^2z/T) = \tilde s^2z^2\,.
    \label{eqS:spinPsi}
    \end{equation}
Here $\psi(x)=\langle x y_\alpha/(1-x y_\alpha)\rangle$ with the angular brackets denoting an average over the distribution $p(y_\alpha)$ of $y_\alpha$, which specifies the mixture composition. Equation~(\ref{eqS:spinPsi}) is to be solved for $T$ as a function of the total density $\rho$. The difference between the two regimes discussed above, i.e.\ outlier and bulk, lies in the way $z=z(\rho,T)$ is determined as we explain next.
    
First, in the outlier regime one has $z=\theta^{-1}$, where $\theta=T/\rho_0 - b$ is the non-zero eigenvalue of $\boldsymbol R_1/M$; notice that $\theta$ has to be negative to give rise to an outlier to the left of the bulk. We can also determine the overlap between the (normalized) instability vector $\boldsymbol v$ and the normalized uniform vector $\hat{\boldsymbol u}=\boldsymbol u/\sqrt{M}$ as
    \begin{equation}
    |\boldsymbol v^\mathsf{T} \hat{\boldsymbol u}|^2 = \max \left \{ -\frac{\tilde s^2}{\theta^2}\left [ 1 + \frac{T^2}{\rho^2\tilde s^2 G'(-\rho\tilde s^2\theta^{-1}/T)} \right ], 0 \right \}
    \label{eqS:evec}        
    \end{equation}
where $G(x)=\langle (x - 1/y_\alpha)^{-1} \rangle$ is the so-called resolvent, the prime denotes derivative and all quantities, i.e.\ $\theta$, $T$ for a given $\rho$, are evaluated on the spinodal curve. Following Ref.~\cite{Sear2003} we shall refer to instabilities with non-vanishing overlap between $\boldsymbol v$ and $\hat{\boldsymbol u}$ as condensation (C), and as demixing otherwise. Equation~({\ref{eqS:evec}}) thus provides information on the nature of the instability and is one of our key results. 

\begin{figure}[hbt]
  \begin{center}  
    \includegraphics[trim={0 0 0 1.0cm},clip,width=\columnwidth]{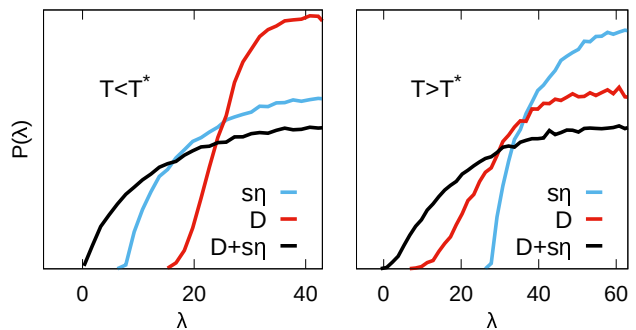}
    \caption{Exemplary realization of eigenvalue distributions of $\boldsymbol D$ and $s\boldsymbol \eta$. The sum $\boldsymbol D+s\boldsymbol \eta$ becomes dominated by the interaction complexity for $T<T^*$, and by the composition $p(y_\alpha)$ for $T>T^*$. We used for $p(y_\alpha)$ a Beta distribution. For visualization purposes the spectra have been shifted to have matching means.}
    \label{fig:spectrum}
  \end{center}  
\end{figure}

Turning next to the bulk regime, we have already written Eq.~\eqref{eqS:evec} in a form that applies also there.
The first argument of the $\max$ function in~\eqref{eqS:evec} is then negative and we have demixing behavior, with the lowest eigenvalue of the bulk $\boldsymbol D+s\boldsymbol \eta$ determining thermodynamic stability. For the restricted case of uniform composition, $\boldsymbol D=(T/\bar\rho)\boldsymbol I$ is proportional to the identity matrix and just shifts the spectrum of $s\boldsymbol\eta$ by $T/\bar\rho$; the entire demixing regime is then described by a linear spinodal line $T\propto \bar\rho$~\cite{Sear2003}. For non-uniform composition, on the other hand, the spread of eigenvalues in $\boldsymbol D$ can dominate at high enough $T$~\cite{Bouchbinder2021} as illustrated in Fig.~\ref{fig:spectrum}. The edge of the bulk of the spectrum is then determined by $\boldsymbol D$ and in the spinodal condition~\eqref{eqS:spinPsi} one has $z=-T/(\tilde s^2 \rho y_{\max})$, where $y_{\max}=\max_\alpha y_\alpha$. The corresponding eigenvector of $\boldsymbol D$ only has a single non-zero entry, and we find as a result that the instability direction $\boldsymbol v$ becomes concentrated on a few species. For low enough $T$, on the other hand, the $s\boldsymbol \eta$ term dominates in $\boldsymbol D+s\boldsymbol \eta$ and $z$ has a larger value maximizing $F(g) = -\tilde s^2g + (1/g)\psi(-\rho \tilde s^2g/T)$, i.e.\ $z=g^*$ with $g^*$ determined from $F'(g^*)=0$. The threshold temperature $T^*$ separating these two cases (see Fig.~\ref{fig:spectrum}) is the one where $F'(-T/\tilde s^2 \rho y_{\max})=0$, and is given explicitly by
    \begin{equation}
    T^* = \tilde s\rho y_{\max}\sqrt{\left \langle  %\frac{y_\alpha^2}{(y_{\max}-y_\alpha)^{2}} 
    y_\alpha^2/(y_{\max}-y_\alpha)^{2}
    \right \rangle }.
    \label{eqS:DIAGtransition}
\end{equation}

Finally, the transition between condensation (outlier) and demixing (bulk) regimes occurs when the two solutions meet each other, namely when $g^*=\theta^{-1}$. Along the spinodal curve this condition defines a threshold density $\rho^*$ such that, for $\rho<\rho^*$, the spinodal is condensation-like and otherwise of demixing type.

Taken together, our results give a complete characterization of the spinodal line for any composition $p(y_\alpha)$ via equation~(\ref{eqS:spinPsi}), and of the nature of the spinodal instability via~(\ref{eqS:evec}). In addition to the condensation-demixing transition~\cite{Sear2003} Eq.~(\ref{eqS:DIAGtransition}) reveals the novel possibility of {\em composition-driven demixing}. We show below that this is not a simple entropic effect, but arises instead from an interplay between the composition and the complexity in the interactions.

\textit{Example 1: Uniform composition.} The case $p(y_\alpha) = \delta(y_\alpha-1)$, where all mixture components have the same density, has been studied in~\cite{Sear2003,Moran2019}. We revisit it briefly in order to illustrate the formalism developed so far. Starting from Eq.~(\ref{eqS:spinPsi}), straightforward algebra yields $T/\rho + z^{-1} + z \tilde s^2 = 0$. In the condensation regime, $z=\theta^{-1}$ with $\theta=T/(1-\rho)-b$, giving a quadratic~\footnote{Sear and Cuesta~\cite{Sear2003} omitted the explicit entropic solvent contribution, resulting in the spinodal curve always being linear in $\rho$. Including the solvent entropy as done in other recent studies~\cite{Jacobs2017,Jacobs2021,shrinivas2021}
makes $\theta$ dependent on $T$ and $\rho$ and gives a nonlinear spinodal curve in the condensation regime.}
equation for the spinodal line $T(\rho)$. In the demixing regime, $T^*$ from Eq.~(\ref{eqS:DIAGtransition}) diverges so that thermodynamic stability is always governed by the interaction complexity $s\boldsymbol \eta$. One then finds from the condition $F'(g^*)=0$ that $\tilde sg^*=1 - T/(\rho \tilde s)$, and inserting $z=g^*$ into the spinodal equation yields $T=2\tilde s\rho$.

To understand the nature of the instabilities we refer to Eq.~(\ref{eqS:evec}). In the uniform distribution case we have $G'(-\rho \tilde s^2\theta^{-1}/T)=-(\tilde s^2\theta^{-1} + T/\rho)^{-2}T^2/\rho^2$ and the term in brackets equals $-\theta$ by the spinodal equation. This yields $|\boldsymbol v^\mathsf{T} \hat{\boldsymbol u}|^2 = \max \{1 - \tilde s^2/\theta^2, 0\}$ so that we are in the condensation regime as long as $-\theta=b-T/(1-\rho)>\tilde s$. This is always the case at low densities, where $T$ and hence $T/(1-\rho)$ vanishes along the spinodal,  provided that $b>\tilde{s}$.
Therefore, at low densities $|\boldsymbol v^\mathsf{T} \hat{\boldsymbol u}|=O(1)$ with densities of all species changing by similar amounts at the spinodal instability. As the total density is increased, $-\theta$ decreases and can approach $\tilde s$; $\boldsymbol v$ and $\hat{\boldsymbol u}$ then become orthogonal, resulting in random demixing (RD), with a delocalized instability vector $\boldsymbol{v}$~\cite{Benaych-Georges2011}. 

\label{subsec:ODS}

\textit{Example 2: One dominant species}. Next, we investigate the case of one single dominant species ($\alpha=1$) with relative concentration $y_1>1$, while all other species have $y_2=(M-y_1)/(M-1)<1$ (see Supplemental Material at\cite{} for $M$-dependent effects). This example will show how tuning the density of a few species can change the nature of the spinodal instability at high densities.

As a consequence of the single distinct entry in $\boldsymbol D$, two possibilities exist for the demixing regime, depending on whether the lowest eigenvalue of  $\boldsymbol D+s\boldsymbol\eta$ is controlled by  $s\boldsymbol\eta$ or by the distinct entry in $\boldsymbol D$. Solving the spinodal equation~(\ref{eqS:spinPsi}) and computing the nature of the instability (See Supplemental Material at~\cite{}) yields three different regimes depending on $y_1$. At low densities, the spinodal is dominated by the average interaction $-b$ and by entropic effects, yielding condensation behavior. Increasing $\rho$ results in a transition to demixing. We find explicitly for the instability direction in the demixing regime $|\boldsymbol v^\mathsf{T} \hat{\boldsymbol u}| = 0$ and
\begin{equation}
    |\boldsymbol v^\mathsf{T} \boldsymbol e_1|^2 = \max\left \{\frac{y_1-2}{y_1-1}, 0\right \}
\end{equation}
where $\boldsymbol e_1=(1,0,0,\dots,0)^\mathsf{T}$ is the direction of the dominant species. The $O(1)$ overlap between $\boldsymbol{v}$ and $\boldsymbol{e}_1$ demonstrates that, whenever $y_2>2$, we have {\em composition-driven demixing} (CD) controlled by the dominant species. If, on the other hand, $y_2<2$ the instability is controlled by $s$ and the mixture will undergo random demixing. The transition from C to CD happens at $\rho^* = 1 - y_1 \tilde s/(b\sqrt{y_1-1}+\tilde s)$ and the CD-spinodal for $\rho>\rho^*$ follows $T = y_1\tilde s \rho/\sqrt{y_1-1}$.

\begin{figure}[htb]
  \begin{center}  
    \includegraphics[width=\columnwidth]{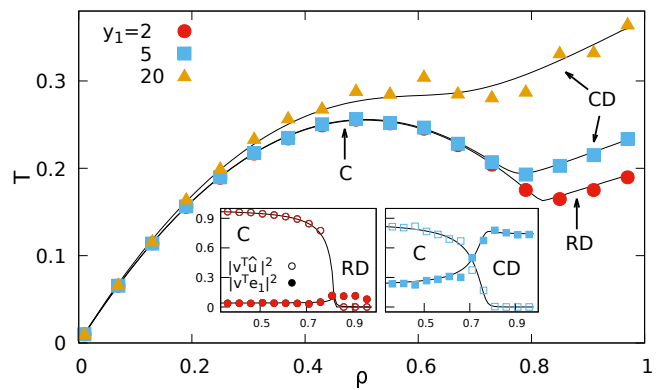}
    \caption{One dominant species case. Spinodal lines for different $y_1=\rho_1/\bar\rho$ at fixed $M=100$, $s=1$ and $b=1$ (lines: theory, symbols: average over $50$ numerical realizations of Hessian matrix). For $y_1>2$ the demixing is controlled by the dominant species. Insets: Projection of instability direction onto the constant vector $\hat{\boldsymbol u}$ and the dominant species direction $\boldsymbol{e}_1$, showing the transitions from condensation (C) to random (RD) and composition-driven demixing (CD), respectively. We keep the $M$-dependence of $\theta$ to account for finite-size effects when comparing to numerics.
    }
    \label{fig:dom1}
  \end{center}  
\end{figure}

In Fig.~\ref{fig:dom1} we compare the predictions for the spinodal curves for different $y_1$ and the corresponding instability direction; the comparison to results from numerical realizations of the Hessian matrix shows excellent agreement. For $y_1>2$ the instability vector is strongly concentrated on the dominant species at high densities, with $v_1/v_2=O(\sqrt{M})$. What is striking in this CD region is the fact that the share of the dominant species in the instability direction is much larger than expected from entropic considerations, which would predict $v_1/v_2 \sim y_1 / y_2 = O(1)$. This strong {\em composition imbalance amplification} is our key insight into instabilities in complex mixtures. It results from an interplay of entropic effects and complexity in the interactions ($\tilde{s}>0$); indeed the CD regime would be absent in the limit $\tilde{s}\to 0$ (where $\rho^*\to 1$).

\begin{figure*}[htb]
  \begin{center}  
    \includegraphics[width=\textwidth]{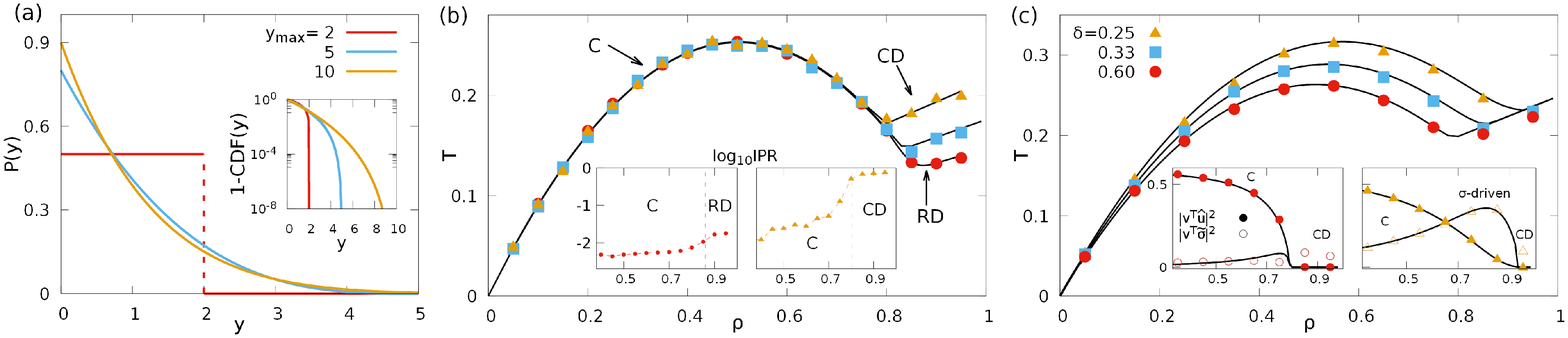}
    \caption{(a) Beta distribution $p(y)$ with $y_{\min}=0$, $r=0$ and $t=y_{\max}-2$. The inset shows the complement of the cumulative distribution function CDF(y), highlighting the different tail behaviors. (b) Spinodal line at fixed $M=600$, $b=1.0$, $s=1.5$ for same distribution parameters (corresponding colors) as in (a). Inset: inverse participation ratio (IPR), showing the localization transition to CD on the right ($y_{\max}=10$). (c) Spinodal lines for Beta distribution with $y_{\max}=5$ and systematic interactions, for different values of $\delta = \tilde s/\sqrt{\textrm{Var}_\sigma + \tilde s^2}$. Inset: projection of instability direction onto the two principal interaction directions, i.e.\ the uniform vector $\hat{\boldsymbol{u}}$ resulting from the volume exclusion and the remaining systematic interaction $\tilde{\boldsymbol{\sigma}}=\boldsymbol{\sigma}-\langle\sigma\rangle\boldsymbol{u}$; for small enough $\delta$ a new regime with instability direction dominated by systematic interactions appears.
    }
    \label{fig:beta}
  \end{center}  
\end{figure*}

\textit{Example 3: Beta distribution}.
We now turn our attention to a continuous distribution $p(y_\alpha)$ and show how its shape affects instabilities. We focus on the Beta distribution defined by
\begin{equation}
    p(y) = Z_{r,t}^{-1}(y-y_{\min})^r(y_{\max}-y)^t, \quad y\in [y_{\min},y_{\max}]
    \label{eq:beta}
\end{equation}
with $Z_{r,t}=(y_{\max}-y_{\min})^{r+t+1}B(r+1,t+1)$ and $B(x,y)$ the Beta function. The requirement $\langle y\rangle=1$ reduces the number of free parameters to three. 

The spinodal equation~\eqref{eqS:spinPsi} can be solved numerically in the condensation (C) regime. More interesting here is the demixing regime. According to Eq.~(\ref{eqS:DIAGtransition}), for temperatures greater than a threshold temperature $T^*$, the demixing instability changes from being determined by the interaction complexity (RD) to being governed by the composition (CD), i.e.\ the distribution of $y$. Inserting~(\ref{eq:beta}) into~(\ref{eqS:DIAGtransition}) one finds
\begin{equation}
    T^* = \tilde s\rho y_{\max}\sqrt{(%\frac{
    Z_{r,t-2}/
    %}{
    Z_{r,t}
    %}%
    )\langle y^2\rangle_{r,t-2}}
    \label{eqS:tcBeta}
\end{equation}
where $\langle \cdot \rangle_{l,m}$ denotes the average over the Beta distribution~\eqref{eq:beta} with exponent parameters $l$ and $m$. For $T>T^*$ the spinodal equation~(\ref{eqS:spinPsi}) can be solved analytically, yielding
\begin{equation}
    T = \tilde s\rho y_{\max}\sqrt{%\frac{
    (Z_{r,t-1}/
    %}{
    Z_{r,t}
    %}
    )\langle y\rangle_{r,t-1}}.
    \label{eqS:spinDIAG}
\end{equation}
Comparing~(\ref{eqS:tcBeta}) and~(\ref{eqS:spinDIAG}), we see that whenever $Z_{r,t-1}\langle y\rangle_{r,t-1} > Z_{r,t-2}\langle y^2\rangle_{r,t-2}$ the demixing spinodal is dictated by the mixture composition. This condition is independent of the total density $\rho$ and controlled only by the shape of $p(y)$. If e.g.\ we fix $y_{\min}=0$, this condition for CD reduces to $t>r+3$, meaning that the upper edge of the distribution has a much longer tail than its lower edge. In other words, the transition occurs whenever a small fraction of species has significantly larger density than the average, mirroring our results from Example 2.

Regarding the nature of the instability, we can exploit the results of Ref.~\cite{Lee2016} to show that the instability direction is delocalized across species in the RD regime ($T<T^*$). In the CD regime, the instability direction is concentrated on a few dominant species that have entries of $O(1)$ in $\boldsymbol{v}$ and will, therefore, dictate the nature of the spinodal instability. Translating the results of~\cite{Lee2016} further to our context, the contribution of the highest-density species (denoted by $\boldsymbol e_1$) to the instability direction is $|\boldsymbol v^\mathsf{T}\boldsymbol e_{1}|^2 = 1 - (T^*/T)^2$, which is independent of $\rho$ along the CD spinodal. For the subsequent dominant species ($j>1$) one has $|\boldsymbol v^\mathsf{T}\boldsymbol e_{j}|^2\sim 1/(M^{\gamma}T^2|\lambda_j-\lambda_1|^2)$ for some exponent $\gamma>0$ that depends on the shape of $p(y)$.

The above results are the continuum analog of the ones obtained in the single dominant species case: by changing the upper edge of the distribution $p(y)$ one can control the nature of instabilities, from delocalized to partially concentrated onto a few dominant species. Fig.~\ref{fig:beta} summarizes this example by showing the complete spinodal line in each regime. To demonstrate the localization of the instability direction onto the few dominant species, we also show the inverse participation ratio (IPR) along the spinodal. The IPR is defined as $\textrm{IPR}=\sum_\alpha v_\alpha^4$ and so expected to be  $O(M^{-1})$ for delocalized instabilities, while it reaches $O(1)$ whenever the instability is concentrated on a few species. 

Here again one observes composition imbalance amplification in the CD regime: if purely entropic effects were at play, the components of the instability direction $\boldsymbol{v}$ should be distributed according to $p(y)$. This would yield a much lower IPR ($O(1/M)$ rather than $O(1)$) than we find, cf.\ Fig.~\ref{fig:beta}. The effect again requires $s>0$, i.e.\ complexity in the underlying interactions.

Finally, we illustrate in Fig.~\ref{fig:beta}.c how the presence of systematic interactions creates further competition between different forms of demixing. To illustrate this, we consider second virial coefficients of the form $\epsilon_{\alpha\gamma}=-b + s\eta_{\alpha\gamma} - \sigma_\alpha\sigma_\gamma$, with an additional systematic term parameterized by an interaction strength $\sigma_\alpha$ associated with each species. We write $\langle \sigma \rangle$ and $\textrm{Var}_\sigma$ for the mean and variance of the $\sigma_\alpha$ and introduce the parameter $\delta = \tilde s/\sqrt{\textrm{Var}_\sigma + \tilde s^2}$ measuring the relative strength of the random and systematic interactions. As density is increased, for sufficiently small $\delta$, the systematic interaction now dominates the instability, cf.\  Fig.~\ref{fig:beta}.c. The inclusion of any finite number of similarly structured interactions into our framework is straightforward and we discuss the extension of existing models~\cite{Graf2021,Carugno2021} in the Suppplemental Material~\cite{}.

To conclude, we have provided a framework for understanding instabilities in mixtures of arbitrary composition, allowing also for the straightforward inclusion of volume exclusion and systematic interactions. In particular we have obtained an exact equation for the spinodal line in the limit of many %mixture 
components, $M\gg 1$, for mixtures with complex interactions as proposed by Sear and Cuesta~\cite{Sear2003}. In simple yet paradigmatic examples we showed that a small number of higher-density mixture components can strongly control the nature of instabilities through a surprising interplay between entropic effects and the complexity of interactions, resulting in a strong composition imbalance amplification. This new form of instability is the main physical insight of the present letter. Since in many biological mixtures different components are present in different amounts, we expect instabilities to phase separation in such systems to be strongly dictated by the components with the highest concentration, strongly influencing the structures and phases found in the steady state. Our results thus also point to a new route for biological systems to control patterns of phase separation by fine-tuning mixture composition imbalances.

\section{Acknowledgments}
This work was supported by the German Research Foundation (DFG) under grant numbers SO 1790/1-1 and KR 3844/5-1.

%\nocite{*}
\bibliographystyle{apsrev4-1}
%\bibliography{refs}

%

\end{document}